\documentstyle[12pt]{article}
\begin{document}

\vspace {.5in}
\begin{center}

{\large {\bf  Generalized Chen-Wu type cosmological model}}

\vspace{.35in}
{\bf Moncy V. John$^{*}$ and  K. Babu Joseph}

Department of Physics, Cochin University of Science and Technology,

Kochi 682022, India.

\vspace{.25in}
{\bf Abstract}

\end{center}

Recent measurements require modifications in conventional cosmology
by way of introducing components other than ordinary matter into the
total energy density in the universe. On the basis of some
dimensional considerations in line with quantum cosmology, Chen and
Wu [W. Chen and Y. Wu, Phys.  Rev. D {\bf 41}, 695 (1990)] have
argued that an additional component, which corresponds to an
effective cosmological constant $\Lambda $ must vary as $a^{-2}$ in
the classical era.  Their decaying-$\Lambda $ model assumes inflation
and yields a value for $q_{0}$, which is not compatible with
observations. We generalize this model by arguing that the Chen-Wu
ansatz is applicable to the total energy density of the universe and
not to $\Lambda $ alone.  The resulting model, which has a coasting
evolution (i.e., $a \propto t$), is devoid of the problems of
horizon, flatness, monopole, cosmological constant, size, age and
generation of density perturbations. However, to avoid serious
contradictions with big bang nucleosynthesis, the model has to make
the predictions $\Omega _{m} = 4/3$ and $\Omega _{\Lambda }=2/3$,
which in turn are at variance with current observational values.

\medskip

{\bf PACS  No(s): 98.80.-k}

\vspace{1.5in}

$^{*}$Permanent address: Department of Physics, St. Thomas College,
Kozhencherri 689641, Kerala, India. e-mail: moncy@stthom.ernet.in

\vspace{.5in}
\newpage

Recent measurements \cite{perl} of the cosmic deceleration parameter,
which point to the need of having some new energy density in the
present universe, in addition to the usual
relativistic/nonrelativistic matter density have caused some
sensation \cite{krauss}. Several other measurements like that of the
combination of the Hubble parameter $H_{0}$ and the  age $t_{0}$ of
the present universe, gravitational lensing, etc., also indicate such
a possibility.  Candidates for such an additional component include
vacuum energy with density $\rho _{\Lambda }$ (identical to that due
to a cosmological constant $\Lambda $, with equation of state
$p_{\Lambda } = -\rho _{\Lambda }$) and "quintessence" \cite{dave} 
with density $\rho _{q}$ (with a general equation of
state $p_{q} = w\; \rho _{q}$; $-1<w<0$ - examples are fundamental
fields and macroscopic objects such as light, tangled cosmic
strings), the former being considered often in the literature.  The
above observations specifically show that if the new component is
$\rho _{\Lambda }$, then its magnitude should be comparable to that
of matter density $\rho _{m}$.  Decaying vacuum cosmologies [4-8]
(and references therein) are phenomenological models, which  conceive
a time-varying $\Lambda $ as an attempt to describe how $\rho
_{\Lambda }$ attains such small values in the present universe.  In
this report, we study one of the pioneering decaying vacuum models
\cite{chen} and suggest an alternative scenario  which is
conceptually more sound.  Though the resulting model faces some
serious problems when concrete theoretical predictions, either on
nucleosynthesis or on the density parameters $\Omega _{m}$ and
$\Omega _{\Lambda }$ are compared with observations, it has  several
positive features and raises certain fundamental issues which invite
serious consideration.

First we recall that Chen and Wu \cite{chen}, while introducing
their
widely discussed   model mentioned above, have made an interesting
argument in favor of an $a^{-2}$ variation of the effective
cosmological constant on the basis of some dimensional
considerations
in line with quantum cosmology. Their reasoning is as follows:
Since there is no other fundamental energy scale available, one can
always write $\rho _{\Lambda}$, the energy density corresponding to
the effective cosmological constant as the Planck density ($\rho
_{pl} = c^{5}/\hbar G^{2} = 5.158\times 10^{94}$  gm cm$^{-3}$ )
times a dimensionless product of quantities.  Assuming that $\rho
_{\Lambda} $ varies as a power of the scale factor $a$, the natural
ansatz is

\begin{equation}
\rho _{\Lambda} \propto \frac {c^{5}}{\hbar G^{2}} \left[ 
\frac {l_{pl}}{a}\right]^{n}, \label{eq:rholambdachen1}
\end{equation}
where $l_{pl} = (\hbar G/c^{3})^{1/2} = 1.616\times 10^{-33}$ cm is
the Planck length. The authors argue that $n=2$ is a preferred
choice. It is easy to verify that $n<2$ (or $n>2$) will lead to a
negative (positive) power of $\hbar $ appearing explicitly on the
right hand side of the above equation. Such an $\hbar $-dependent
$\rho _{\Lambda }$ would be quite unnatural in the classical Einstein
equation for cosmology, much later than the Planck time.  However, it
shall be noted that $n=2$ is just right to survive the semiclassical
limit $\hbar \rightarrow 0$. This choice is further substantiated by
noting that $n \leq 1$ or $n\geq 3$ would lead to a value of $\rho
_{\Lambda }$ which violates all observational bounds. Thus the
Chen-Wu ansatz is

\begin{equation}
\rho _{\Lambda} = \frac {\gamma }{8\pi G a^{2}},
\label{eq:rholambdachen2} 
\end{equation}
where $\gamma $ is a phenomenological constant parameter. (Here
onwards we set $\hbar = c= k_{B} = 1$, except when stating explicit
results).  Assuming that only the total energy-momentum is conserved,
they obtain, for the relativistic era,

\begin{equation}
\rho _{r} = \frac {A_{1}}{a^{4}} + \frac {\gamma }{8\pi G
a^{2}} \equiv \rho _{r}^{cons.} + \rho
_{r}^{noncons.}\label{eq:rhorchen}
\end{equation}
and for the nonrelativistic era,

\begin{equation}
\rho _{nr} = \frac {A_{2}}{a^{3}} + \frac {2\gamma }{8\pi G a^{2}}
\equiv \rho _{nr}^{cons.} + \rho _{nr}^{noncons.},
\label{eq:rhonrchen}
\end{equation}
where $A_{1}$ and $A_{2} $ are to be positive. The Chen-Wu model thus
differs from the standard model in that it has a decaying
cosmological constant and that the matter density has conserving and
nonconserving parts [given by the first and second terms respectively
in the right hand sides of Eqs. (\ref{eq:rhorchen}) and
(\ref{eq:rhonrchen})]. By choosing $\gamma $ appropriately, they hope
to arrange $\rho _{\Lambda} $ and the nonconserving parts in $\rho
_{r}$ and $\rho _{nr}$ to be insignificant in the early universe so
that the standard model results like nucleosynthesis are undisturbed.
But for the late universe, it can have many positive features like
providing the missing energy density in the flat and inflationary
models, etc.. The model predicts creation of matter, but the authors
argue that the creation rate is small enough so that it is
inaccessible to observations.

The important criticisms one can raise in this regard are the
following: Conversely to the requirement that the conserving part of
matter density dominate the early universe (for the standard model
results to remain undisturbed), one can deduce that in their model,
the standard model results are applicable  only to the same   part of
matter density.  The nonconserving parts are, in fact, created almost
entirely in the late universe. But the abundance of light nuclei etc.
are verified for the present universe and this implies that the
conserving part is still substantial.  This in turn will create some
problem with observations.  For example, let us assume that  the
present era is nonrelativistic  and  $\rho _{nr}^{cons.}$ is at least
equal to $\rho _{nr}^{noncons.}$.  Since the vacuum density is only
one-half the latter quantity [See Eqs.  (\ref{eq:rholambdachen2}) and
(\ref{eq:rhonrchen})], for a $k=0$ universe in which $\Omega _{m}+
\Omega _{\Lambda } =1$, the deceleration parameter at present will
be $q_{0} = (\Omega _{m}/2)-\Omega _{\Lambda }= 0.2$.  This is not
compatible with the observations mentioned earlier \cite {perl}.

Also, since it is conceived that their model is not different from
the  standard model in the early universe, to avoid the cosmological
problems, they have to assume the occurrence of inflation, which in
turn is driven by the vacuum energy.  But they apply their ansatz
only to the late-time vacuum energy density (which corresponds to the
cosmological constant) and not to that during inflation. The stress
energy associated with the vacuum energy is identical to that of a
cosmological constant and it is not clear how they distinguish them
while applying the ansatz.

Lastly, it can genuinely be asked whether $\rho _{\Lambda }$ is the
only quantity to which the Chen-Wu ansatz  be applied. An equation
analogous to (\ref{eq:rholambdachen1}) can be written for any kind
of
energy density by using a similar reasoning and it can be argued
that
$n=2$ is a preferred choice for each one of them in the late
universe.  Certainly, this will bring in some fundamental issues
which need serious consideration, but there is  a priori no reason
to
forbid such an investigation.

In this report, we present a cosmological model by applying the
Chen-Wu ansatz to the total energy density $\tilde {\rho}$ of the
universe, in place of the vacuum density alone. If the Chen-Wu
argument is valid for $\rho _{\Lambda }$, then it should be valid
for
$\tilde {\rho }$ too.  In fact, this ansatz is better suited to
$\tilde {\rho}$ rather than to $\rho _{\Lambda }$, since the Planck
era is characterized by the Planck density for the universe, above
which quantum gravity effects become important. Hence we modify
the
ansatz to write

\begin{equation}
\tilde{\rho } =A \frac {c^{5}}{\hbar G^{2}}\left[ \frac
{l_{pl}}{a}\right] ^{n}, \label{eq:rhotilde1}
\end{equation}
where $A$ is a positive dimensionless  constant.  As indicated
above,
when $\tilde {\rho}$ is the sum of various components and each
component is assumed to vary as a power of the scale factor $a$,
then
the Chen-Wu argument can be applied  to conclude that $n=2$ is a
preferred choice for each component. Violating this will force the
inclusion of $\hbar $ -dependent terms in $\tilde {\rho }$, which
would look unnatural in a classical theory.  Not only for the Chen
and Wu model, in all of FRW cosmology, this argument may be used to
forbid the inclusion of substantial energy densities which do not
vary as $a^{-2}$ in the classical epoch.

At first sight, this may appear as a grave negative result. But let
us face it squarely and proceed to the next logical step of
investigating the implications of an $a^{-2}$ variation of $\tilde
{\rho}$.  If the total pressure in the universe is denoted as
$\tilde
{p}$, then the above result that the conserved quantity $\tilde
{\rho
}$ in the FRW model varies as $a^{-2}$ implies $\tilde {\rho } + 3
\tilde {p} =0$. This will lead to a coasting cosmology (i.e.,
$a\propto t$). Components with such an equation of state are known
to
be strings or textures \cite{kamion}. Though such models are
considered in the literature, it would be unrealistic to consider
the
present universe as string-dominated. A crucial observation which
makes our model with $ \tilde {\rho }$ varying as $ a^{-2}$
realistic
is that this variation leads to string-domination only if we assume
$\tilde {\rho }$ to be unicomponent. Instead, if we assume, as done
in inflationary, Chen and Wu and many other models
(Friedmann-Lamaitre-Robertson-Walker cosmologies) that $\tilde
{\rho
}$ consists of parts corresponding to relativistic/ nonrelativistic
matter (with equation of state $p_{m} = w \rho _{m} $ where $w=
1/3$
for relativistic and $w=0$ for nonrelativistic  cases) and also to
a
time-varying cosmological constant (with equation of state
$p_{\Lambda} = -\rho_ {\Lambda }$),  i.e., if we assume,

\begin{equation}
\tilde {\rho } = \rho _{m} + \rho _{\Lambda }, \qquad \tilde {p} =
p_{m} + p_{\Lambda }, \label{eq:rhoptilde}
\end{equation}
then the condition $\tilde {\rho } + 3\tilde {p} =0$ will give

\begin{equation}
\frac {\rho _{m}}{\rho _{\Lambda }} = \frac
{2}{1+3w}\label{eq:m/l}.
\end{equation}
In other words, the modified Chen-Wu ansatz leads to the conclusion
that if the universe contains matter and vacuum energies, then
vacuum
energy density should be comparable to matter density.  This, of
course, will again lead to a coasting cosmology, but this time a
realistic one.  (The Ozer-Taha model \cite{ozer} in its
relativistic
era and the models in \cite{lopez,jj} are approximately some such
models, but they start from  different sets of assumptions.)

$\rho _{m}$ or $\rho _{\Lambda }$, which varies as $a^{-2}$, may
sometimes be mistaken for  strings but it should be noted that the
equations of state we assumed for these quantities are different
from
that for strings and are what they ought to be to correspond to
matter density and vacuum energy density respectively.  It is true
that components with equations of state $p=w\; \rho $  should obey
$
\rho \propto a^{-3(1+w)}$, but this is valid when those components
are separately conserved. In our case, we have only assumed that
the
total energy density is conserved and not the parts corresponding
to
$\rho _{m}$ and $\rho _{\Lambda } $ separately. Hence, as in the
Chen-Wu model, there can be creation of matter from vacuum, but we
shall show later in this report that again the present creation
rate
is too small to make any observational consequences.

The solution to the Einstein equations in an FRW model with $\tilde
{\rho } + 3\tilde {p}=0$, for all the three cases $k=0, \pm 1$, is
the coasting evolution

\begin{equation}
a(t) = m t, \label{eq:a}
\end{equation}
where $m$ is some proportionality constant. The total energy
density
is then

\begin{equation}
\tilde {\rho } = \frac {3}{ 8\pi G} \frac {(m^{2} +k)}{a^{2}}.
\label{eq:rhotilde2} 
\end{equation}
Comparing this with (\ref{eq:rhotilde1}) (with $n=2$), we get
$m^{2}
+k = 8\pi A/3$.  We shall now show that this simple picture of the
universe is devoid of many of the
cosmological problems encountered in the standard model.

First let us consider the horizon problem. A necessary condition
for
the solution of this problem is \cite{hu} $a(t_{s}) \int
_{t_{pl}}^{t_{s}} dt/a(t)> [a(t_{s})/a(t_{0})] H_{0}^{-1}$, where
$t_{s}$ is the time by which the horizon problem is solved.  Using
our expression (\ref{eq:a}) for $a(t)$, this condition gives $t_{s}
\geq e\; t_{pl}$.  Thus shortly after the Planck era, the horizon
problem is solved in this model.  Since causality is established at
such early times, the monopole problem will also disappear.

The predictions regarding the age of the universe in the model is
obvious from Eq. (\ref{eq:a}). Irrespective of the value of $m$, we
get the combination $H_{0}t_{0}$ as equal to unity, which is well
within the bounds. Thus there is no age problem in this model. We
can
legitimately define the critical density as $\rho _{c} \equiv
(3/8\pi
G) (\dot {a}^{2}/a^{2})$, so that Eq. (\ref{eq:rhotilde2}) gives

\begin{equation}
\tilde {\Omega } \equiv \frac {\tilde {\rho }}{\rho _{c}} = \left[
1-
\frac 
{3k}{8\pi A }\right] ^{-1}. \label{eq:omegatilde}
\end{equation}

As in the standard model, we have $\tilde {\Omega } = 1$ for $k=0$
and $\tilde {\Omega }>1 $ ($\tilde{\Omega }<1$) for $k= +1$
($k=-1$).
But unlike the standard model, $\tilde {\Omega }$ is a constant in
time.  This is not surprising; in an FRW model with total energy
density $\tilde {\rho }$, one can always write the time-time
component of Einstein equation in the form

\begin{equation}
\tilde {\Omega }-1 = \left[ \frac {8\pi G}{3} \frac {\tilde {\rho
}a^{2}}{k} -1 \right] ^{-1}.
\end{equation}
When $\tilde {\rho }$ varies  $a^{-3}$ or $a^{-4}$, the flatness
problem appears and the reason can be understood from this
equation.
But in the present case, since $\tilde {\rho }$ varies as $a^{-2}$,
$\tilde {\Omega }$ will remain a constant.  Using Eqs.
(\ref{eq:rhoptilde}) and (\ref{eq:m/l}), we get

\begin{equation}
\Omega _{m} \equiv \frac {\rho _{m}}{\rho _{c}} = \frac {2\tilde
{\Omega }}{3(1+w)}, \qquad \Omega _{\Lambda } \equiv \frac {\rho
_{\Lambda }}{\rho _{c}} = \frac {(1+3w) \tilde {\Omega }}{3(1+w)}.
\label{eq:omegaml} 
\end{equation}
For the matter dominated era, the predictions are $\Omega _{m} =
2\tilde {\Omega }/3$ and $\Omega _{\Lambda } = \tilde {\Omega }/3$.
Note that also the density parameter $\Omega _{m}$ is
time-independent  and hence there is no flatness problem in this
model. As mentioned above, the  model predicts that the energy
density corresponding to the cosmological constant is comparable
with
matter density and this solves the cosmological constant problem
too.
It can also be seen that according to the model, the observed
universe, characterised by the present Hubble radius has a size
equal
to the Planck length at the end of Planck epoch and this indicates
that the problem with the size of the universe does not appear
here.
For the investigation of other problems, we have to study the
thermal
evolution of the universe as envisaged in the model.

In the early relativistic era,  temperature $T$ is associated with
the relativistic matter density $\rho _{r}$ as $\rho _{r} = (\pi
^{2}/30) N(T) T^{4}$, where $N(T)$ is the effective number of spin
degrees of freedom at temperature T. In the present model,

\begin{equation}
\rho _{r} = \frac {3\tilde {\Omega }}{8\pi G} \frac {1}{ (\sqrt
{2}t)^{2}}. \label{eq:rhorp}
\end{equation}
This gives

\begin{equation}
T=\left[ \frac {3}{8\pi G}\frac {30 \tilde {\Omega }}{\pi ^{2}N}
\right]^{1/4} \frac {1}{(\sqrt {2} t)^{1/2}}.
\end{equation}

These expressions may be compared with the corresponding
expressions
in the standard model:

\begin{equation}
\rho _{s.m.} = \frac {3}{8\pi G} \frac {1}{ (2t)^{2}},
\label{eq:rhosm}
\end{equation}

\begin{equation}
T_{s.m.}=\left[ \frac {3}{8\pi G}\frac {30 }{\pi ^{2}N}
\right]^{1/4} \frac {1}{(2 t)^{1/2}}.
\end{equation}
Considering the fact that according to observation $\tilde {\Omega
}^{1/4}$ is close to unity, it can be seen that the values of $\rho
_{r}$ and $T$ attained at time $t$ in the standard model are
attained
at time $\sqrt {2} t$ in the present model. Thus the thermal
history
in the present model can be expected to be nearly the same as
that in the standard model.  But the time-dependence of the scale
factor is different in our model and this helps  to solve the
cosmological problems.

So far we have considered $\tilde {\Omega }$ to be a free parameter,
related by Eq. (\ref {eq:omegatilde}) to the constant $A$, which in
turn is to be understood to come from some deep quantum cosmological
theory.  An interesting way to estimate the constant $\tilde {\Omega
}$ is to consider the implications of the model for nucleosynthesis
\cite{kolb}. From (\ref{eq:rhorp}) and (\ref{eq:rhosm}), one can
deduce that the Hubble parameter in the present model is related to
that in the standard model according to $H=\sqrt {2/\tilde {\Omega }}
\; H_{s.m.}$. This modifies the ratio of interaction rate to Hubble
parameter as $\Gamma /H = \sqrt {\tilde {\Omega }/2} \; \; \Gamma
/H_{s.m.}$.  To avoid any variation of the freezing temperature with
that in the successful standard model, one has to accept a value
$\tilde {\Omega } \approx 2$. This leads us to the predictions $
\Omega _{m} \approx 4/3 $ and $ \Omega _{\Lambda
}\approx 2/3$, which are in contradiction with the recent
measurements \cite{perl} since the corresponding point is outside the
error ellipses in the $\Omega _{m}-\Omega _{\Lambda }$ plot. This
discrepancy with observation is a serious problem which requires
detailed analysis and refinement in the model.

The possibility of the generation of density perturbations on
scales
well above the present Hubble radius, in the interval between the
Planck time $t_{pl}$ and  the time of decoupling $t_{dec}$ can be
studied  by evaluating the communication distance light can travel
between these two  times \cite{liddle}. In the present model,
$d_{comm} (t_{pl}, t_{dec}) = a_{0} \int _{t_{pl}}^{t_{dec}}
dt/a(t)
= 0.627 \times 10^{6} \hbox {Mpc}$, where we have used $t_{dec}
\approx 10^{13}$ s,  the same as that in the standard model.  Thus
the coasting  evolution in this case has the communication distance
between $t_{pl}$ and $t_{dec}$ much larger than the present Hubble
radius ($\approx 4000 $ Mpc) and hence it can generate density
perturbations on scales of that order. It is interesting to note
that
Liddle \cite {liddle} has precluded coasting evolution as a viable
means to produce such perturbations and argued that only inflation
($\ddot {a} >0$) can perform this task, thus "closing the
loopholes"
in the arguments of Hu {\sl et. al.} \cite {hu}. But it is
worthwhile
to point out that his observations are true only for a  model which
coasts from $t_{pl}$ to $t_{nuc}$ (where $ t_{nuc} \approx 1$ s  is
the time of nucleosynthesis) and thereafter evolves according to
the
standard model. In our case, the evolution is coasting throughout
the
history of the universe and hence his objection is not valid.

A bonus point of the present approach, when compared to all the
other
aforementioned models may now be noted. In those  models, the
communication distance between $t_{nuc}$ and $t_{dec}$, or for that
matter the communication distance from any time after the
production
of particles (assuming this to occur at the end of inflation) to
the
time $t_{dec}$ will be only around $200 h^{-1}$ Mpc, $0.6<h<0.8$
\cite{liddle}.  Thus density perturbations on scales above the
present Hubble radius cannot be generated in them in the period
when
matter is present.  This is because inflation cannot enhance the
communication distance after it. The only means to generate the
observed density perturbations is then to resort to quantum
fluctuations of the inflaton field. The present model is at a more
advantageous position than the inflationary models in this regard
since the communication distance between $t_{nuc}$ and $t_{dec}$ in
this case is $d_{comm} (t_{nuc}, t_{dec}) = a_{p} \int
_{t_{nuc}}^{t_{dec}} dt/a(t) = 1.45 \times 10^{5} \hbox {Mpc}$,
which
is much greater than the present Hubble radius. So we can consider
the generation of the observed density perturbations as a late-time
classical behavior too.

Lastly we check the rate of matter creation in the model.  Assuming
the present universe to be dominated by nonrelativistic matter, we
can calculate the rate of creation per unit volume as $a^{-2}
d(\rho
_{m}a^{3})/dt \mid _{p} = \rho _{m0}H_{0}$.  This creation rate is
only one-third of that in the steady state model. Creation of
matter
or radiation with an average rate given above will be inaccessible
to
test  and does not pose a serious objection to the model.

It was recently argued \cite{dave} that a smooth time-varying
$\Lambda $ is ill defined and unstable and that the only valid way of
introducing an additional energy component is to replace $\Lambda $
with a fluctuating, inhomogeneous component.  (Such an energy
component is the quintessence, mentioned in the introduction.)
Notwithstanding this and other serious problems with observations
(either the big bang nucleosynthesis or the prediction of density
parameters), it is worth  noting that if we take quantum cosmology
seriously, generalizing the Chen-Wu ansatz is a logical conclusion
and that it leads to a realistic cosmological scenario, which does
not have many of the  problems in the standard model,
including that of the generation of density perturbations  in the
late classical  epoch itself.

We acknowledge the valuable comments by the unknown referee, with
thanks. MVJ is grateful to IUCAA, Pune for its hospitality, where
part of this work was done.

\end{document}